\documentclass{article}
\usepackage{spconf,amsmath,graphicx,hyperref}

\title{Exploring Resolution-Wise Shared Attention in Hybrid Mamba-U-Nets for Improved Cross-Corpus Speech Enhancement}
\name{Nikolai Lund Kühne$^{\star}$ \qquad Jesper Jensen$^{\star\dagger}$ \qquad Jan Østergaard$^{\star}$ \qquad Zheng-Hua Tan$^{\star}$}
 \address{$^{\star}$Department of Electronic Systems, Aalborg University, 9220, Denmark. \\
$^{\dagger}$Oticon A/S, 2765, Denmark}
\usepackage{balance}                % Balances the columns of the last page
\usepackage{dirtytalk} %benytte \say til at lave pæne anførselstegn
\usepackage{siunitx}						% Flot og konsistent praesentation af tal og enheder med \si{enhed} og \SI{tal}{enhed}
\usepackage[official]{eurosym} 

\usepackage{soul}
\usepackage{makecell}

\makeatletter
\newcommand*{\bigs}[1]{{\hbox{$\left#1\vbox to10\p@{}\right.\n@space$}}}
\makeatother

\makeatletter
\newcommand*{\biggs}[1]{{\hbox{$\left#1\vbox to17\p@{}\right.\n@space$}}}
\makeatother

\usepackage{blindtext}

\usepackage{tikz}  %Pakke til plot af funktioner og grafer
\usetikzlibrary{decorations.pathreplacing,calligraphy}
\usepackage[framemethod=tikz]{mdframed}
\usetikzlibrary{shapes.geometric, arrows, arrows.meta}
\usepackage{subcaption}
\usepackage{graphicx}
\usepackage{float}
\usepackage{booktabs}
\usepackage{multirow}
\usepackage{dblfloatfix}
\DeclareSIUnit[quantity-product = ]\percent{\char`\%}
\usepackage{comment}
\usetikzlibrary{decorations.text}
\usetikzlibrary{shapes.geometric, arrows}
\usetikzlibrary{positioning}
\usepackage{adjustbox}

\usetikzlibrary{intersections}
\usepackage{cite}
\usepackage{amsmath,amssymb,amsfonts}
\usepackage{algorithmic}
\usepackage{graphicx,color}
\usepackage{textcomp}
\usepackage{xcolor}
\usepackage{hyperref}
\hypersetup{hidelinks=true}
\usepackage{algorithm,algorithmic}
\def\BibTeX{{\rm B\kern-.05em{\sc i\kern-.025em b}\kern-.08em
    T\kern-.1667em\lower.7ex\hbox{E}\kern-.125emX}}

\AtBeginDocument{\definecolor{tmlcncolor}{cmyk}{0.93,0.59,0.15,0.02}\definecolor{NavyBlue}{RGB}{0,86,125}}

\usepackage{bm}
\usepackage{comment}

\usepackage{pifont} % Checkmark
\makeatletter
\newcommand*\bigcdot{\mathpalette\bigcdot@{.5}}
\newcommand*\bigcdot@[2]{\mathbin{\vcenter{\hbox{\scalebox{#2}{$\m@th#1\bullet$}}}}}
\makeatother
\usepackage{tikz}  %Pakke til plot af funktioner og grafer
\usepackage{url}

\definecolor{mygreen}{RGB}{127,201,127}
\definecolor{mypurple}{RGB}{190,174,212}
\definecolor{myorange}{RGB}{253,192,134}
\definecolor{myblue}{RGB}{72,213,210}

\usepackage{balance}
\usepackage{nicematrix}
\begin{document}
\ninept

\maketitle

\begin{abstract}
Recent advances in speech enhancement have shown that models combining Mamba and attention mechanisms yield superior cross-corpus generalization performance. At the same time, integrating Mamba in a U-Net structure has yielded state-of-the-art enhancement performance, while reducing both model size and computational complexity. Inspired by these insights, we propose RWSA-MambaUNet, a novel and efficient hybrid model combining Mamba and multi-head attention in a U-Net structure for improved cross-corpus performance. Resolution-wise shared attention (RWSA) refers to layerwise attention-sharing across corresponding time- and frequency resolutions. Our best-performing RWSA-MambaUNet model achieves state-of-the-art generalization performance on two out-of-domain test sets. Notably, our smallest model surpasses all baselines on the out-of-domain DNS 2020 test set in terms of PESQ, SSNR, and ESTOI, and on the out-of-domain EARS-WHAM\_v2 test set in terms of SSNR, ESTOI, and SI-SDR, while using less than half the model parameters and a fraction of the FLOPs.
\end{abstract}
\section{Introduction}
Speech enhancement aims at removing background noise from speech signals, thereby improving speech intelligibility and quality. It has a wide range of applications such as hearing assistive devices, mobile communication systems, and speaker verification.

In the past decade, research in deep-learning based speech enhancement has included a wide range of neural architectures \cite{xie2025survey}. The applied architectures include convolutional neural networks \cite{fu2016snr,kolbaek2020loss}, diffusion models \cite{lu2022conditional,richter2023speech} and generative adversarial networks (GANs) \cite{michelsanti17_interspeech,fu2019metricgan}. Recently, attention based neural architectures such as Transformers and Conformers have been the most prevalent, as these models have demonstrated state-of-the-art (SOTA) performance on multiple benchmarks \cite{mp-senet, lu2023explicit}. However, multi-head attention (MHA) based models scale quadratically with input size in terms of computational complexity \cite{gu2024mamba}.
This led to a newfound interest in recurrent models that instead scale linearly with respect to input size. Two of such models, Mamba \cite{gu2024mamba} and Extended Long Short-Term Memory (xLSTM) \cite{xlstm}, have already demonstrated SOTA in-domain speech enhancement performance \cite{semamba, wang2025mamba, kuhne25_interspeech}. Moreover, recent works such as Mamba-SEUNet \cite{wang2025mamba} and MUSE \cite{lin24h_interspeech} have demonstrated the effectiveness of U-Nets in speech enhancement, offering similar or improved enhancement performance with fewer model parameters and lower computational complexity. However, the efficacy of U-Nets for cross-corpus speech enhancement has not been explored yet.

Cross-corpus generalization performance is important for speech enhancement systems, as such systems may be expected to operate across a diverse range of acoustic conditions, and it is infeasible to include all recording conditions, noise types, and speakers in the training dataset. Unfortunately, sequence models like LSTM, xLSTM, and Mamba have demonstrated worse generalization performance compared to purely attention based models \cite{pandey2020cross, pandey2022self, 11359486}. As an alternative to purely attention based models, the hybrid MambAttention model \cite{11359486} was recently proposed. By combining Mamba with a shared time- and frequency-MHA module, MambAttention achieved SOTA cross-corpus generalization performance on two out-of-domain test sets across all evaluation metrics used.

Based on the SOTA cross-corpus generalization of MambAttention, we hypothesize that explicitly aligning global time- and frequency relations is critical for robust cross-corpus speech enhancement. To realize this hypothesis, we propose resolution-wise shared attention (RWSA), which is shared layerwise time- and frequency-MHA modules across corresponding time- and frequency resolutions in Mamba-UNets. Our proposed RWSA-MambaUNet employs MambAttention blocks, which have demonstrated superior generalization performance \cite{11359486}. By introducing RWSA, our best-performing RWSA-MambaUNet model achieves SOTA generalization performance on two out-of-domain test sets with different speakers, noise types, and recording conditions across PESQ, SSNR, ESTOI, and SI-SDR, at a significantly lower computational complexity than the baselines. Remarkably, even our smallest model outperforms all baselines on most metrics for cross-corpus generalization with less than half the model parameters. Code is publicly available.\footnote{\url{https://github.com/NikolaiKyhne/RWSAMamba-UNet}}
Our major contributions are summarized as follows:
\begin{itemize}
    \item We propose RWSA-MambaUNet, a novel and efficient hybrid model using resolution-wise shared attention in a U-Net archictecture for improved cross-corpus generalization.
    \item We demonstrate that RWSA is essential for the SOTA cross-corpus performance of our RWSA-MambaUNet models.
    \item Our best-performing model surpasses existing SOTA baselines on two out-of-domain test sets across all evaluation metrics used, while requiring significantly fewer FLOPs.
\end{itemize}

\begin{figure*}[thb]
    \centering
    \input{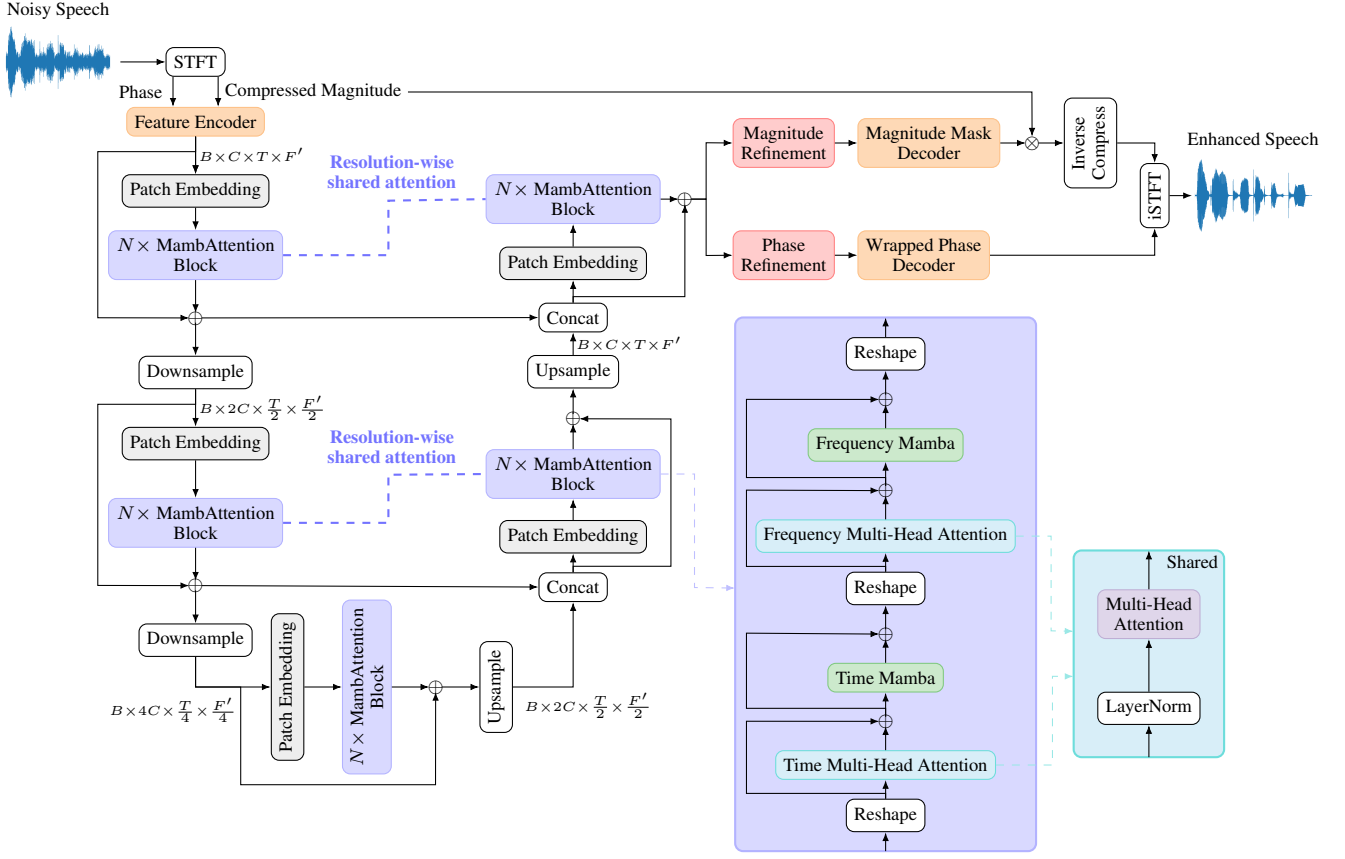}
    \caption{Overall structure of our proposed RWSA-MambaUNet. Resolution-wise shared attention (purple dashed lines) is layerwise sharing of MHA modules within MambAttention blocks across corresponding resolutions between the upsampling and downsampling path. To simplify the figure, we have not depicted the residual connections between the output of the feature encoder, and the outputs of both refinement layers.}
    \label{fig:RWSAMambaUNet}
\end{figure*}
\section{Method}
\subsection{MambAttention}
Our RWSA-MambaUNet consists of multiple MambAttention blocks \cite{11359486} at different time- and frequency resolutions. We employ these blocks, as they have demonstrated SOTA speech enhancement generalization performance \cite{11359486}. MambAttention blocks comprise bidirectional Mamba blocks across time (T-Mamba) and frequency (F-Mamba), as well as shared time-MHA (T-MHA) and frequency-MHA (F-MHA) modules. Given an input $\bm{X}\in\mathbb{R}^{B\times C\times T\times F}$, where $B$ is the batch size, $C$ is the number of channels, and $T$ and $F$ denote the number of time frames and frequency bins, respectively, the forward pass of a MambAttention block is given by \cite{11359486}:
\begin{align}
    \bm{X}_{\mathrm{Time}} &= \mathrm{reshape}(\bm{X},[B\cdot F,T,C]), \\
    \bm{X}_1 &= \bm{X}_{\mathrm{Time}} + \text{T-MHA}(\mathrm{LN}(\bm{X}_{\mathrm{Time}})), \\
    \bm{X}_2 &= \bm{X}_1 + \text{T-Mamba}(\bm{X}_1), \\
    \bm{X}_{\mathrm{Freq.}} &= \mathrm{reshape}(\bm{X}_2,[B\cdot T,F,C]), \\
    \bm{X}_3 &= \bm{X}_{\mathrm{Freq.}} + \text{F-MHA}(\mathrm{LN}(\bm{X}_{\mathrm{Freq.}})),\\
    \bm{X}_4 &= \bm{X}_3 + \text{F-Mamba}(\bm{X}_3), \\
    \bm{Y} &= \mathrm{reshape}(\bm{X}_4,[B, C, T,F]),
\end{align}
where $\mathrm{reshape}(\mathrm{input}, \mathrm{size})$ reshapes the input to a desired size, and LN is Layer Normalization. MambAttention utilizes the T- and F-Mamba blocks from SEMamba \cite{semamba}, and the output $\bm{X}_\mathrm{out}$ of each T- and F-Mamba block is given by:
\begin{multline}
    \bm{X}_\mathrm{out} = \mathrm{Conv1D}(\mathrm{Concat}(\mathrm{Mamba}(\bm{X}_\mathrm{in}), \\\mathrm{flip}(\mathrm{Mamba}(\mathrm{flip}(\bm{X}_\mathrm{in}))))),
\end{multline}
where $\bm{X}_\mathrm{in}$ is the input to the T- and F-Mamba blocks, and $\mathrm{Mamba}(\cdot)$, $\mathrm{flip}(\cdot)$, $\mathrm{Concat}(\cdot)$, and $\mathrm{Conv1D}(\cdot)$ is the unidirectional Mamba, sequence flipping, concatenation, and 1D transposed convolution. 

\subsection{Model overview}
 Our RWSA-MambaUNet is portrayed in \autoref{fig:RWSAMambaUNet}.

\textbf{Preprocessing and feature encoder:} Before the noisy speech waveform $\bm{y}\in\mathbb{R}^L$ is processed by the feature encoder, a complex spectrogram is computed through a short-time Fourier transform (STFT). The input $\bm{Y}_{in}\in\mathbb{R}^{T\times F\times 2}$ to the feature encoder then becomes the compressed magnitude spectrum $(\bm{Y}_m)^c\in\mathbb{R}^{T\times F}$, extracted via power-law compression \cite{powerlawcompression}, concatenated with the wrapped phase spectrum $\bm{Y}_p\in\mathbb{R}^{T\times F}$. The feature encoder is identical to the one used in MP-SENet \cite{mp-senet} and thus increases the number of input channels from $2$ to $C$ and halves the frequency dimension from $F$ to $F'=F/2$. It consists of two convolution blocks, each comprising a 2D convolutional layer, an instance normalization, and a PReLU activation, sandwiching a dilated DenseNet \cite{densenet}.

\textbf{U-Net architecture:} The output of the feature encoder is processed by multiple MambAttention blocks with skip
connections, patch embedding layers, as well as convolutional downsampling and upsampling blocks, following a U-Net-style architecture. The patch embeddings, originally proposed in \cite{qiu2023mb}, consist of depthwise separable and deformable convolutions, which facilitates learning intricate fine-grained acoustic details. The MambAttention blocks focus on capturing time and frequency dependencies across acoustic features at different time- and frequency resolutions. The final magnitude and phase refinement layers comprise a patch embedding, $N$ stacks of TF-Mamba blocks from \cite{semamba}, and a $3\times 3$ convolution. 

\textbf{Resolution-wise shared attention:} We believe attention-sharing across corresponding resolutions aids reconstruction in the U-Net, since global time- and frequency relationships are aligned at matching resolution levels across both the down- and upsampling paths. RWSA (purple dashed lines in \autoref{fig:RWSAMambaUNet}) leverages the fact that distinct MambAttention blocks are used at corresponding time- and frequency resolutions in both paths of the U-Net. By sharing the T- and F-MHA modules not only within each individual MambAttention layer but also across layers in both the downsampling and upsampling paths, the model jointly aligns global temporal- and spectral dependencies across multiple resolution scales. As we will demonstrate, RWSA improves generalization performance, while minimizing the memory cost of the attention blocks.

\textbf{Magnitude mask and wrapped phase decoder:} Finally, after the magnitude and phase refinement layers, the clean magnitude and phase spectra are estimated through the magnitude mask and wrapped phase decoder, respectively. Following \cite{lu2023explicit}, both the magnitude mask and wrapped phase decoder consist of a dilated DenseNet, followed by a sub-pixel convolution \cite{shi2016real}, an instance normalization, and a PReLU activation. In the magnitude mask decoder, this
is followed by a deconvolution block reducing the output channels from $C$ to 1. A learnable sigmoid function with $\beta = 2$ is used to estimate the magnitude mask, as in \cite{mp-senet}. In the wrapped phase decoder, the sub-pixel convolution is followed by two parallel 2D convolutional layers yielding the pseudo-real and pseudo-imaginary part components. The
clean wrapped phase spectrum is estimated using the two-argument arctangent function \cite{mp-senet}.
The final enhanced waveform is recovered by applying an inverse STFT to the estimated clean magnitude spectrum and estimated wrapped phase spectrum.

We follow MambAttention \cite{11359486} and use a linear combination of loss functions, including a PESQ-based GAN discriminator, along with time, magnitude, complex, phase, and consistency losses.

\section{Experiments}
\begin{table*}[h]
    \centering
    \caption{In-domain and out-of-domain speech enhancement performance. Models are trained on  VB-DemandEx. All baselines are trained using their originally provided code. Best reported mean is marked bold.}
    \setlength{\tabcolsep}{2.5pt}

\begin{adjustbox}{width=1\textwidth}
\begin{tabular}{@{}lcr|cccc|cccc|cccc@{}}
    \toprule
        \multirow{3}{*}{Dataset} & & &\multicolumn{4}{c|}{In-Domain} & \multicolumn{8}{c}{\: \: \: Out-Of-Domain} \\ \cmidrule(lr){4-7}\cmidrule(l){8-15}
        & & &\multicolumn{4}{c|}{\textit{VB-DemandEx}} & \multicolumn{4}{c|}{\textit{DNS 2020}} & \multicolumn{4}{c}{\textit{EARS-WHAM\_v2}}\\ \hline
    Model &Params & FLOPs & PESQ         & SSNR                & ESTOI       & SI-SDR          &PESQ         & SSNR                & ESTOI       & SI-SDR & PESQ         & SSNR                & ESTOI       & SI-SDR   \\ \hline
    Noisy & - & - & 1.625   & -1.068   & 0.630   & 4.976      & 1.582    & 6.218    & 0.810 & 9.071 & {1.235} & {-0.803} & {0.640} & {5.359}  \\ \hline
    xLSTM-SENet \cite{kuhne25_interspeech} & 2.20M & 80.71G&  $2.973\scriptscriptstyle\pm 0.051$   & $7.933\scriptscriptstyle\pm 0.133$             & $0.795\scriptscriptstyle\pm 0.008$   & $16.414\scriptscriptstyle\pm 0.317$       & $1.724\scriptscriptstyle\pm 0.368$    & $3.246\scriptscriptstyle\pm 1.332$ &  $0.686\scriptscriptstyle\pm 0.097$ & $3.412\scriptscriptstyle\pm 3.482$  & {$1.505\scriptscriptstyle\pm 0.151$} & {$0.446\scriptscriptstyle\pm 0.566$} & {$0.559\scriptscriptstyle\pm 0.053$} & {$1.396\scriptscriptstyle\pm 2.141$} \\
    LSTM-SENet \cite{kuhne25_interspeech}  & 2.34M &88.59G &$3.002\scriptscriptstyle\pm 0.026$ & $\bf{7.981\scriptscriptstyle\pm 0.210}$      &  $0.802\scriptscriptstyle\pm 0.003$ &  $16.637\scriptscriptstyle\pm 0.123$ & $1.984\scriptscriptstyle\pm 0.454$ & $4.901\scriptscriptstyle\pm 1.656$ & $0.724\scriptscriptstyle\pm 0.117$ & $4.749\scriptscriptstyle\pm 3.346$  & {$1.570\scriptscriptstyle\pm 0.179$} & {$0.854\scriptscriptstyle\pm 0.773$} & {$0.566\scriptscriptstyle\pm 0.083$} & {$1.916\scriptscriptstyle\pm 2.894$}  \\
    SEMamba \cite{semamba}& 2.25M & 65.46G&$3.002\scriptscriptstyle\pm 0.022$ & $7.590\scriptscriptstyle\pm 0.177$      & $0.800\scriptscriptstyle\pm 0.003$ &$16.593\scriptscriptstyle\pm 0.159$       & $2.281\scriptscriptstyle\pm 0.134$  & $5.837\scriptscriptstyle\pm 1.033$ & $0.820\scriptscriptstyle\pm 0.028$ & $9.298\scriptscriptstyle\pm 1.576$  &  {$1.631\scriptscriptstyle\pm 0.053$}  & {$0.921\scriptscriptstyle\pm 0.508$} & {$0.603\scriptscriptstyle\pm 0.026$} & {$2.809\scriptscriptstyle\pm 0.523$} \\
    MP-SENet \cite{mp-senet}& 2.05M  & 74.29G &$2.935\scriptscriptstyle\pm 0.065$ & $7.641\scriptscriptstyle\pm 0.283$ & $0.787\scriptscriptstyle\pm 0.010$ & $16.202\scriptscriptstyle\pm 0.318$       & $2.666\scriptscriptstyle\pm 0.010$ & $7.369\scriptscriptstyle\pm 0.382$ & $0.875\scriptscriptstyle\pm 0.009$ & $13.665\scriptscriptstyle\pm 0.892$  & {$1.862\scriptscriptstyle\pm 0.097$} & {$2.107\scriptscriptstyle\pm 0.270$} & {$0.677\scriptscriptstyle\pm 0.029$} & {$6.090\scriptscriptstyle\pm 0.672$}  \\
    MambAttention \cite{11359486}& 2.33M & 65.52G& $\bf{3.026\scriptscriptstyle\pm 0.007}$ & $7.674\scriptscriptstyle\pm 0.411$      &  $\bf{0.801\scriptscriptstyle\pm 0.002}$ &  $\bf{16.684\scriptscriptstyle\pm 0.095}$  &  $2.919\scriptscriptstyle\pm 0.118$ & $8.133\scriptscriptstyle\pm 0.733$ & $0.911\scriptscriptstyle\pm 0.009$ & $15.169\scriptscriptstyle\pm 1.363$ & {$2.010\scriptscriptstyle\pm 0.053$} & {$2.505\scriptscriptstyle\pm 0.224$} & {$0.725\scriptscriptstyle\pm 0.020$} & {$7.348\scriptscriptstyle\pm 0.445$} \\
    \hline
RWSA-MambaUNet-XS& \bf{1.02M} & \bf{9.22G}& $2.893\scriptscriptstyle\pm 0.009$ & $7.041\scriptscriptstyle\pm 0.073$      &  $0.780\scriptscriptstyle\pm 0.002$ &  $15.212\scriptscriptstyle\pm 0.064$  &  $2.940\scriptscriptstyle\pm 0.019$ & $9.421\scriptscriptstyle\pm 0.132$ & $0.922\scriptscriptstyle\pm 0.002$ & $14.722\scriptscriptstyle\pm 0.120$ & $1.987\scriptscriptstyle\pm 0.023$ & $3.106\scriptscriptstyle\pm 0.188$ & $0.729\scriptscriptstyle\pm 0.006$ & $8.541\scriptscriptstyle\pm 0.347$  \\
RWSA-MambaUNet-S& 1.95M & 14.91G& $2.936\scriptscriptstyle\pm 0.006$ & $7.350\scriptscriptstyle\pm 0.013$      &  $0.789\scriptscriptstyle\pm 0.002$ &  $15.453\scriptscriptstyle\pm 0.065$  &  $3.042\scriptscriptstyle\pm 0.020$ & $9.670\scriptscriptstyle\pm 0.024$ & $0.930\scriptscriptstyle\pm 0.001$ & $15.047\scriptscriptstyle\pm 0.079$ & $2.033\scriptscriptstyle\pm 0.030$ & $3.334\scriptscriptstyle\pm 0.069$ & $0.740\scriptscriptstyle\pm 0.008$ & $8.946\scriptscriptstyle\pm 0.297$  \\
RWSA-MambaUNet-M& 3.91M & 28.47G & $3.001\scriptscriptstyle\pm 0.006$ & $7.490\scriptscriptstyle\pm 0.113$      &  $0.800\scriptscriptstyle\pm 0.002$ &  $16.017\scriptscriptstyle\pm 0.085$  &  $\bf{3.126\scriptscriptstyle\pm 0.011}$ & $\bf{10.019\scriptscriptstyle\pm 0.074}$ & $\bf{0.936\scriptscriptstyle\pm 0.001}$ & $\bf{15.600\scriptscriptstyle\pm 0.065}$ & $\bf{2.101\scriptscriptstyle\pm 0.0011}$ & $\bf{3.690\scriptscriptstyle\pm 0.054}$ & $\bf{0.763\scriptscriptstyle\pm 0.005}$ & $\bf{9.198\scriptscriptstyle\pm 0.250}$  \\
    \bottomrule
\end{tabular}
\end{adjustbox}

    \label{tab:generalization}
\end{table*}
\begin{table*}[h]
    \centering
    \caption{Ablation study. Default configurations for our RWSA-MambaUNet is with RWSA, and with T- and F-MHA modules in the MambAttention blocks. Models are trained on VB-DemandEx. Best reported mean is marked bold.}
    \setlength{\tabcolsep}{2.5pt}
\begin{adjustbox}{width=1\textwidth}
\begin{tabular}{@{}lcc|cccc|cccc|cccc@{}}
    \toprule
        \multirow{3}{*}{Dataset} & & &\multicolumn{4}{c|}{In-Domain} & \multicolumn{8}{c}{\: \: \: Out-Of-Domain} \\ \cmidrule(lr){4-7}\cmidrule(l){8-15}
        & & &\multicolumn{4}{c|}{\textit{VB-DemandEx}} & \multicolumn{4}{c|}{\textit{DNS 2020}} & \multicolumn{4}{c}{\textit{EARS-WHAM\_v2}}\\ \hline
    Model &Params & FLOPs & PESQ         & SSNR                & ESTOI       & SI-SDR          &PESQ         & SSNR                & ESTOI       & SI-SDR & PESQ         & SSNR                & ESTOI       & SI-SDR   \\ \hline
    Noisy & - & - & 1.625   & -1.068   & 0.630   & 4.976      & 1.582    & 6.218    & 0.810 & 9.071 & {1.235} & {-0.803} & {0.640} & {5.359}   \\ \hline
RWSA-MambaUNet-S& 1.95M & 14.91G& $\bf{2.936\scriptscriptstyle\pm 0.006}$ & $\bf{7.350\scriptscriptstyle\pm 0.013}$      &  $\bf{0.789\scriptscriptstyle\pm 0.002}$ &  ${15.453\scriptscriptstyle\pm 0.065}$  &  $\bf{3.042\scriptscriptstyle\pm 0.020}$ & $\bf{9.670\scriptscriptstyle\pm 0.024}$ & $\bf{0.930\scriptscriptstyle\pm 0.001}$ & $\bf{15.047\scriptscriptstyle\pm 0.079}$ & $\bf{2.033\scriptscriptstyle\pm 0.030}$ & $\bf{3.334\scriptscriptstyle\pm 0.069}$ & $\bf{0.740\scriptscriptstyle\pm 0.008}$ & $\bf{8.946\scriptscriptstyle\pm 0.297}$  \\
w/o RWSA & 1.98M & 14.91G& $2.906\scriptscriptstyle\pm 0.017$ & $7.119\scriptscriptstyle\pm 0.004$      &  $0.782\scriptscriptstyle\pm 0.002$ &  $15.275\scriptscriptstyle\pm 0.124$  &  $2.956\scriptscriptstyle\pm 0.026$ & $9.461\scriptscriptstyle\pm 0.030$ & $0.924\scriptscriptstyle\pm 0.001$ & $14.838\scriptscriptstyle\pm 0.210$ & $1.957\scriptscriptstyle\pm 0.031$ & $3.010\scriptscriptstyle\pm 0.097$ & $0.731\scriptscriptstyle\pm 0.003$ & $8.448\scriptscriptstyle\pm 0.161$  \\
w/o MHA modules \cite{wang2025mamba}& \bf{1.88M} & \bf{14.45G}& $2.915\scriptscriptstyle\pm 0.021$ & $7.162\scriptscriptstyle\pm 0.091$      &  $0.786\scriptscriptstyle\pm 0.003$ &  $\bf{15.456\scriptscriptstyle\pm 0.116}$  &  $2.932\scriptscriptstyle\pm 0.009$ & $9.308\scriptscriptstyle\pm 0.109$ & $0.922\scriptscriptstyle\pm 0.001$ & $14.757\scriptscriptstyle\pm 0.035$ & $1.922\scriptscriptstyle\pm 0.024$ & $3.096\scriptscriptstyle\pm 0.069$ & $0.714\scriptscriptstyle\pm 0.010$ & $7.901\scriptscriptstyle\pm 0.280$  \\
    \bottomrule
\end{tabular}
\end{adjustbox}

    \label{tab:ablations}
\end{table*}
\subsection{Datasets}
We train and evaluate our models on the VB-DemandEx dataset \cite{11359486}. The dataset contains 10,840 noisy-clean pairs of audio clips for training, 730 for validation, and 840 for testing. The clean speech originates from the VoiceBank corpus \cite{voicebank}, where 26 distinct speakers are used for training, 2 distinct speakers are used for validation, and 2 distinct speakers are used for testing. The noisy audio clips are created by mixing clean samples with noise from the DEMAND database \cite{demand} as well as babble and speech-shaped noise at 7 segmental SNRs (SSNRs) ([$-10, -5, 0, 5, 10, 15, 20$] dB).

In addition, we train and evaluate our models on the large-scale Deep Noise Suppression Challenge 2020 dataset (DNS 2020) \cite{dns2020}. DNS 2020 contains 500 hours of clean speech from 2,150 speakers and more than 180 hours of noise clips. Since DNS 2020 has no validation set, we use the validation set generated in \cite{11359486}. Using the official script provided in \cite{dns2020}, we generate 3,000 hours of noisy-clean pairs of audio clips for training with SSNRs uniformly sampled between \SI{-5}{dB} and \SI{15}{dB}. This yields \SI{1.08}{M} 10-second audio clips. We use the DNS 2020 test set without reverberation for evaluating our models. The test set contains 150 noisy-clean pairs, generated from audio clips spoken by 20 distinct speakers.

Finally, we also evaluate cross-corpus performance on the \SI{16}{kHz} version of the EARS-WHAM\_v2 test set \cite{richter24_interspeech, wichern19_interspeech}. EARS-WHAM\_v2 contains clean speech, recorded in an anechoic chamber, from 107 distinct speakers. The clean speech covers reading tasks in 7 reading styles, emotional reading, conversational speech, and freeform speech. Using the script provided in \cite{richter24_interspeech}, we mix the clean speech from speakers \textit{p102} to \textit{p107} with noise recordings from the WHAM! dataset \cite{wichern19_interspeech} at SNRs randomly sampled in the interval [-2.5, 17.5] dB. This results in 886 noisy-clean pairs for testing. 
\subsection{Implementation details} Unless otherwise stated, all experimental details and training configurations match those presented in MambAttention \cite{11359486}.
We train on 30,600 point audio segments, and use an FFT order of 510, a Hann window size of 510, and a hop size of 120 for all STFTs. Moreover, we use a magnitude spectrum compression
factor of $c = 0.3$. In the MambAttention blocks, we use $h=8$ attention heads for the bottleneck layers and $h=4$ attention heads anywhere else. Checkpoints are saved every 250 steps, and for evaluation we select the checkpoint that obtains the highest PESQ score on the validation set. 
Models trained on VB-DemandEx and DNS
2020 are trained for \SI{550}{k} and \SI{950}{k} steps respectively, with
a batch size $B=8$ on four NVIDIA L40S GPUs.
\autoref{tab:hyperparams} provides important hyperparameters for our proposed RWSA-MambaUNet models.
\begin{table}[H]
    \centering
    \caption{Model hyperparameters for the proposed RWSA-MambaUNet models.}
\begin{adjustbox}{max width=\textwidth}
\begin{tabular}{@{}lccc@{}} 
\toprule
Model &\makecell{\# Channels\\ $C$} & \makecell{\# Blocks\\ $N$} & Params \\
\midrule
RWSA-MambaUNet-XS & 16 & 2 & 1.02M\\
RWSA-MambaUNet-S & 16 & 4 & 1.95M\\
RWSA-MambaUNet-M & 24 & 4 & 3.91M\\
\bottomrule
\end{tabular}
\end{adjustbox}
    \label{tab:hyperparams}
\end{table}
\subsection{Evaluation metrics}
We apply wide-band PESQ \cite{pesq} to evaluate the speech quality of the enhanced speech. Moreover, we report the waveform-matching-based evaluation metrics SSNR \cite{ssnr} and
scale-invariant signal-to-distortion ratio (SI-SDR) \cite{le2019sdr}. The intelligibility of the enhanced speech is predicted using extended short-time objective intelligibility (ESTOI) \cite{jensen2016algorithm}. Across these measures, higher
values indicate better performance. Finally, we report FLOPs, which are calculated based on processing a single audio sample on one GPU.
We train all models with 5 different seeds, and report the mean and standard deviation.

\section{Results}
\subsection{Generalization performance}
We evaluate in-domain performance on the VB-DemandEx dataset. 
For assessing cross-corpus generalization performance, we evaluate on two out-of-domain test sets with different noise, speaker, and recording conditions from DNS 2020 \cite{dns2020} and EARS-WHAM\_v2 \cite{richter24_interspeech, wichern19_interspeech}. For simplicity, we rename the LSTM baseline from \cite{kuhne25_interspeech} to LSTM-SENet.

In \autoref{tab:generalization}, we report in- and out-of-domain speech enhancement performance. From \autoref{tab:generalization}, it is clear that our RWSA-MambaUNet-XS outperforms all the LSTM-SENet, xLSTM-SENet, SEMamba, MP-SENet, and the MambAttention baselines on the out-of-domain DNS 2020 test set across PESQ, SSNR, and ESTOI, and on the EARS-WHAM\_v2 test set across SSNR, ESTOI, and SI-SDR, with only \SI{1.02}{M} parameters and \SI{9.22}{G} FLOPs. By doubling the number of layers from 2 to 4, our RWSA-MambaUNet-S further improves both in- and out-of-domain enhancement performance across all metrics. Finally, as shown in \autoref{tab:generalization}, by increasing the number of channels from 16 to 24, our RWSA-MambaUNet-M outperforms all baselines across all used evaluations metrics on both out-of-domain test sets. While bigger in parameter count, RWSA-MambaUNet-M still requires significantly less FLOPs compared to the baselines. We observed no performance gains by further increasing the model size.

Interestingly, compared to the baselines, we only observe consistent SI-SDR improvements on the out-of-domain EARS-WHAM\_v2 test set, which is the only dataset used, where the clean references are recorded in an anechoic chamber \cite{richter24_interspeech}. In comparison, our RWSA-MambaUNet models slightly underperform across the SI-SDR metric on the the in-domain VB-DemandEx and out-of-domain DNS 2020 test sets. We attribute the performance differences across datasets to the characteristics of the reference signals. This behaviour aligns with the findings of \cite{jepsen2025study}, indicating that our RWSA-MambaUNet models primarily learn to reconstruct clean speech.

\subsection{Ablation study}
To understand the performance impact of key aspects of our RWSA-MambaUNet models, we conduct an ablation study on the RWSA and the shared T- and F-MHA modules in the MambAttention blocks, which are used inside our RWSA-MambaUNet models. Since ablations in \cite{11359486} already demonstrated the positive performance impact of sharing the parameters of the T- and F-MHA modules in the MambAttention blocks, we omit this ablation.

The ablation study in \autoref{tab:ablations} reveals that removing RWSA from the RWSA-MambaUNet-S model decreases both cross-corpus generalization performance and in-domain performance while slightly increasing model size. Removing the MHA modules from the MambAttention blocks reduces our RWSA-MambaUNet model to Mamba-SEUNet \cite{wang2025mamba}. From \autoref{tab:ablations}, we observe that removing the MHA modules negatively affects generalization performance, as all metrics across both out-of-domain test sets decrease. 
\subsection{Results on DNS 2020}
To investigate the scalability of our proposed RWSA-MambaUNet models with respect to training dataset size and diversity, we train them on the large-scale DNS 2020 dataset.

\autoref{tab:dns2020} reveals that our RWSA-MambaUNet-XS matches or outperforms the xLSTM-SENet, SEMamba, and MP-SENet baselines on the ESTOI metric, at less than half the parameter count. Moreover, our RWSA-MambaUNet-S  slightly outperforms all baselines except the SOTA MambAttention model on the PESQ and ESTOI metric, while delivering a similar SSNR score to SEMamba.
Finally, our RWSA-MambaUNet-M matches or outperforms all baselines across SSNR and ESTOI, with a lower computational complexity. MambAttention remains slightly superior for in-domain speech enhancement performance as shown in \autoref{tab:generalization} and \autoref{tab:dns2020}.
\begin{table}[H]
    \centering
    \caption{Speech Enhancement performance on DNS 2020. All baselines are trained using their originally provided code. Best reported mean is marked bold.}
    \setlength{\tabcolsep}{2.5pt}

\begin{adjustbox}{width=1\columnwidth}
\begin{NiceTabular}{@{}lcr|cccc@{}} 
\toprule
Model & Params &FLOPs& PESQ & SSNR & ESTOI & SI-SDR  \\
\midrule
Noisy & - & - & $1.582$ & $6.218$ & $0.810$ & $9.071$ \\
\midrule
xLSTM-SENet \cite{kuhne25_interspeech} & $2.20$M & 80.71G& $3.588\scriptstyle\pm 0.017$ & $14.526\scriptstyle\pm 0.482$ & $0.954\scriptstyle\pm 0.001$ & $20.854\scriptstyle\pm 0.226$ \\
LSTM-SENet \cite{kuhne25_interspeech} & $2.34$M & 88.59G& $3.598\scriptstyle\pm 0.031$ & $15.021\scriptstyle\pm 0.168$ & $0.956\scriptstyle\pm 0.002$ & $21.003\scriptstyle\pm 0.215$ \\
SEMamba \cite{semamba} & $2.25$M & 65.46G& $3.594\scriptstyle\pm 0.012$ & $14.830\scriptstyle\pm 0.473$ & $0.955\scriptstyle\pm 0.001$ & $21.035\scriptstyle\pm 0.123$ \\
MP-SENet \cite{mp-senet} & $2.05$M & 74.29G & $3.605\scriptstyle\pm 0.021$ & $14.967\scriptstyle\pm 0.044$ & $0.954\scriptstyle\pm 0.000$ & $20.919\scriptstyle\pm0.021$  \\
MambAttention \cite{11359486} & $2.33$M & 65.52G & $\bf{3.671\scriptstyle\pm0.008}$ & $15.116\scriptstyle\pm0.049$ & $\bf{0.959\scriptstyle\pm0.000}$ & $\bf{21.234\scriptstyle\pm0.033}$ \\
\midrule
RWSA-MambaUNet-XS& \bf{1.02M} & \bf{9.22G} & $3.563\scriptstyle\pm0.002$ & $14.685\scriptstyle\pm0.039$ & $0.955\scriptstyle\pm0.000$ & $20.457\scriptstyle\pm0.016$\\
RWSA-MambaUNet-S & 1.95M & 14.91G & $3.614\scriptstyle\pm0.009$ & $14.869\scriptstyle\pm0.091$ & $0.957\scriptstyle\pm0.000$ & $20.798\scriptstyle\pm0.049$\\
RWSA-MambaUNet-M & 3.91M & 28.47G & $3.649\scriptstyle\pm0.017$ & $\bf{15.119\scriptstyle\pm0.069}$ & $\bf{0.959\scriptstyle\pm0.000}$ & $21.119\scriptstyle\pm0.094$\\
\bottomrule
\end{NiceTabular}
\end{adjustbox}
    \label{tab:dns2020}
\end{table}
\section{Conclusion}
In this paper, we proposed a novel and efficient hybrid RWSA-MambaUNet model for improved cross-corpus speech enhancement performance. Experiments revealed that our best-performing RWSA-MambaUNet model significantly outperforms existing baselines on two very different out-of-domain test sets across all evaluation metrics. Notably, even our smallest RWSA-MambaUNet model outperforms existing state-of-the-art models across most metrics on two out-of-domain corpora, while using significantly fewer parameters and FLOPs. Finally, we demonstrated that resolution-wise shared attention contributes to the superior cross-corpus enhancement performance of our RWSA-MambaUNet models.
% Finally, training and evaluating on the large-scale DNS 2020 dataset reveals that RWSA-MambaUNet is competitive with or outperforms existing state-of-the-art models across several metrics while requiring significantly less FLOPs
% \balance
\bibliographystyle{IEEEbib}
\bibliography{strings, refs}

\vfill\pagebreak

\end{document}